# Kinetics of motile solitons in fluid nematics

Satoshi Aya*,#, Fumito Araoka*

Physicochemical Soft Matter Research Team, RIKEN Center for Emergent Matter Science (CEMS), Japan

**Solitary waves, dubbed "solitons", are special types of waves that propagate for an infinite distance under ideal conditions. These waves are ubiquitously found in nature such as typhoon or neuron signals. Yet, their artificial generation and the control of their propagation remain outstanding challenges in materials science owing to an insufficient understanding of the experimental conditions and theoretical aspects. Herein, a generic strategy for forming particle-like solitons and controlling their kinetics in nematic fluid media is reported. The key to the realisation of the generation of solitons and the control of their kinetics is the coupling between the fluid elasticity and the background flow flux, as evidenced by experimental observations and theoretical approaches. The findings of this study enable the exploration of solitons in a wide range of materials and have technological ramifications for the lossless transport of energy or structures.**

Solitons are packets of spatially localised moving waves that convey energy with very little or no mass transport. Since the first observations of the solitonic behaviour of water waves by Russell in 1834 and the introduction of the term of "soliton" by Zabusky and Kruskal[1], solitons have been identified in various natural phenomena. However, scientists

* *To whom correspondence should be addressed, E-mail: satoshi.aya@riken.jp, fumito.araoka@riken.jp*

# *Current contact address: South China University of Technology, satoshiaya@scut.edu.cn*

have little access to artificial solitons, despite many dedicated theoretical and computational studies. From a mathematical perspective, solitons are recognised by their particle-like nature such as nondissipative and elastic collisional properties as a result of balancing their dispersive effects with the nonlinear counterparts described by a Korteweg-de-Vries or nonlinear Schrödinger equation[2]. This nature of solitons enables the distortion-free long-distance transport of waves or structures, which has attracted considerable interest from both fundamental and technological viewpoints in many branches of science, including atmospheric circulation[3,4], fluid[5] and mechanical waves[6-8], living organisms[9,10], Bose–Einstein condensates[11-13], light confinement, and propagation[14-16].

Very recently, there have been several groundbreaking studies observing solitonic analogues for physical systems such as spintronic[17-19] and fluid systems[20-24]. In most of these cases, the solitons possess topological singularities in spin-like vector fields that are distinct from the far-field background. In other words, the structures of the solitons are topologically protected. Although it is expected that such phenomena are generally triggered by the spontaneous breaking of either the rotational or translational symmetry or both, the primary unsolved questions lie in the understanding of the physics underpinning both the generation and transport of solitons in real materials.

In this article, we describe many unknown dynamic solitonic phenomena closely involved in many systems with a particle-like nature that can be directly generated and even controlled in nematic fluids. A swarm of a novel type of motile solitons is stabilised and "fueled" by the coupling of the dielectric and hydrodynamic convectional responses to low-voltage electric fields. This discovery establishes the fundamental physics and strategy of manipulating the kinetics of solitons in an experimental setting and permits the observation of their time evolution, facilitating further experimental and theoretical studies of dynamic soliton excitations in soft-matter systems.

**Soliton creation**

As a soliton medium, we prepare thin films of frustrative mixtures comprising two nematics, E7 and 4'-butyl-4-heptyl-bicyclohexyl-4-carbonitrile (CCN47) (Fig. 1a). We uniformly align them between two glass plates with the average molecular orientation, denoted by the director $\overline{n}$, parallel to the plates. To create solitons, a sinusoidal or rectangular AC electric field is applied normal to the plates (Fig. 1a). The electric response significantly changes because of the frustration of the dielectric response and the hydrodynamic instability incurred by changing the combination of the dielectric anisotropy, $\Delta\varepsilon = \varepsilon_\parallel - \varepsilon_\perp$, and conductivity anisotropy, $\Delta\sigma = \sigma_\parallel - \sigma_\perp$ (the subscripts $\parallel$ and $\perp$ indicate the directions parallel and perpendicular to $\overline{n}$, respectively). Generally, the combination of dielectricity and conductivity can be classified using de Gennes' notation as $(\Delta\varepsilon\Delta\sigma)$[25]. As E7 and CCN47 exhibit positive $(++)$ and negative $(--)$ anisotropies, respectively, mixing E7 with CCN47 enables smooth sign tuning from $(--)$ to $(++)$ via $(-+)$. As the mixture of anisotropies corresponds to specific parameter strengths, the effects of this tuning can be investigated.

In Figs. 1b–d, we show time-dependent microscopic texture images acquired during soliton creation in a mixture containing 16 wt% E7 for a 10-V electric field at 12 Hz at the specific times designated in Fig. 1e. Two processes—nucleation and packing—are clearly visible.

**Nucleation**. Independent cruciform solitons are randomly excited in space and time (Fig. 1b) as a result of the localisation of ions when an electric field is applied, while they disappear when the field is absent (Supplementary section 1 and 2, Supplementary Figs. S1 and S2). These solitons fluctuate around and remain relatively near their nucleation positions with a space-filling ratio of $X = \pi R^2/A < 0.6$, where $R$ and $A$ are the soliton radius and the total area of the region, respectively. The time dependence of $X$ is fitted by the Kolmogorov–Johnson–Mehl–Avrami (KJMA) expression, which might give insights to the procedures of the migration and the localisation of the ions during the soliton generation (Supplementary Fig. S3), $\ln[\ln 1/\{1 - X\}] = m \ln t + \ln K$, where $X$, $t$, $K$, and $m$ are the space occupancy of the director-deformed region, the time, the temperature-dependent Avrami coefficient, and the exponent, respectively[26]. The value obtained for $m$ is 2.9, suggesting the involvement of two-dimensional homogeneous nucleation and growth processes triggered by the localisation of ions[27].

**Packing.** Centered rectangular packing of the solitons occurs when the filling ratio $X > 0.6$ (Fig. 1d, Supplementary Fig. S4), a value consistent with the close-packed ratio of perfectly hexagonally packed solid spheres in two dimensions (0.604). For such packing states, the solitons are fusion-resistant. Similar packing structures occur in some self-driven systems[24,28,29]

**Structure of solitons.** We investigate the director structure of the solitons. Since the solitons are dynamic with time, magnified views of solitons obtained with an exposure time of 200 ms under crossed polarisers and with a standard tint waveplate (retardation = 530 nm) taken by a conventional camera (Figs. 1f and g) just reflect a time-averaged structure, an array of concentric director looking like patterns with one *s*=+1 topological defect is clear in the midplane of the nematic film. In order to elucidate the dynamic structure, we perform high-frame-rate polarising microscopy at 100 fps. Figure 1h shows a time-dependent oscillation of the structure of the solitons. Surprisingly, unlike the static structure observed in Figs. 1f and g, the director field of the solitons transforms between two defect-less structures, i.e., biconvex and biconcave structures, synchronised with the applied electric field. It is noted that, by analysing the director fields, the biconvex structure has a larger deviation angle of the director in the midplane of the nematic film than the biconcave structure (Supplementary Fig. S5). This explains why the time-averaged structure looks like a concentric domain with a *s*=+1 defect.

As the soliton filling ratio increases, mutual repulsion, separating the solitons by distances of about twice the soliton radius, is observed. As a result of this repulsion, the solitons become resistant to the external stress without coalescence, confirming the pseudoparticle nature of the solitons.

**Conditions of soliton creation: topological state diagram.** The topological pattern that includes the solitons varies depending on the voltage and frequency (Figs. 2a–e). On the basis of the director deformations and flow profiles (Figs. 2f–j), we classify the frustrated states into five categories: (I) ground planar alignment (G-state), (II) ∥-roll pattern (∥-state), (III) rectangular pattern (R-state), (IV) soliton state (Soliton-state), and (V) ⊥-roll pattern (⊥-state). Fig. 2k shows a state diagram for a mixture containing 16 wt% E7, which exhibits $(-+)$ anisotropy.

In the G-state, neither elastic deformation nor hydrodynamic effects occur; therefore, the initial uniform uniaxial alignment is maintained. In the ∥-state, which appears in the lower-frequency region below 9 Hz, only an in-plane splay elastic deformation of the director occurs because of the coupling with the flexion-induced polarisation, the so-called flexoelectric effect[30]. Because the strong surface anchoring effect forces fix the surface directors in the initial state, twist deformation along the electric field direction ($z$ direction) is induced, yielding a one-dimensional periodic pattern. Particle tracking confirms the absence of flow (see Methods). An increase in the voltage in the ∥-state causes a field-induced hydrodynamic convective flow parallel to the initial director ($\overline{n}_0$). The coupling between the flow and the flexoelectric effect induces a two-dimensional version of the ∥-state, i.e., the R-state. The ⊥-state occurs in a dielectric regime in which ions are effectively immobile, but the nematic director oscillates with the applied electric field. The convective flow field parallel to $\overline{n}_0$ arising from an electrohydrodynamic effect (Carr–Helfrich instability) yields a stripe pattern in which the director has a wavy modulation in space, as observed in nematics with $(-+)$ anisotropy[26]. The appearances of the respective roll patterns in the ∥- and ⊥-states significantly differ in terms of the roll direction: rolls in the ∥-state run parallel to the director; those in the ⊥-state are perpendicular. The Soliton-state occurs in the frequency range between the ∥-/R- and ⊥-states and spans a conductive regime in which ionic convection dominates the dielectric oscillation of the director[31]. Accordingly, the soliton system has features of both the ∥ and ⊥-states with a compresence of elastic and convective effects, as discussed below.

We next turn to the major focus of this study: how the controlled anisotropies of dielectricity and conductivity affect the soliton stabilisation mechanism. To systematically examine the effects of the material parameters, we either altered the concentration of E7 or added excesses of the ionic species TBABE to the CCN47. Fig. 2l shows a generalised scheme of the procedural conditions. Solitons are observed only in $(-+)$ mixtures at low electric fields ($\leqq 10$ V), a typical condition in which the Carr–Helfrich electrohydrodynamic mechanism, where a convective flow occurs parallel to the background director, becomes active[25]. In contrast, no observable flow occurs for $(--)$ and $(++)$ mixtures, with the flexoelectric effect appearing for the former and Freederikz-effect-driven uniform reorientation appearing for the latter. Although the present system does not realise $(+-)$ mixtures, the Freederikz effect dominates $(+-)$ mixtures in conventional systems. The strength of the conductivity is also vital in determining the soliton stability. Fig. 2m shows that solitons appear in the intermediate strength range $8 \times 10^{-9} < \sigma < 4 \times 10^{-8}\, \Omega^{-1} \cdot m^{-1}$, where the localisation of ions is realised. This once again confirms that the Carr–Helfrich electrohydrodynamic mechanism is necessary for the presence of solitons (Supplementary Information 8).

**Soliton kinetics**

The soliton system described herein exhibits two types of kinetic regimes— swimming and proliferation—which appear in ascending order as the frequency of the Soliton-state increases at a fixed voltage. Worth noting is that voltage or frequency contributes differently to the kinetics of the solitons as discussed below.

**Swimming regime: collision and elastic reflection.** An increase in the voltage or frequency in the oscillation regime (the transition from the orange to purple-blue regions in Supplementary Fig. S6) produces steady directional motion of the solitons in which the initial cruciform configuration changes to a bug-eye-like pattern because of asymmetric elastic deformation (Figs. 3a and b). The direction of this motion can be tuned by changing either the amplitude or frequency of the electric field. Figs. 3c–h show the time-dependent trajectories of solitons under the control of the tuned amplitude of the electric field at a characteristic frequency of 20 Hz, where the solitons are most stable in their shape. At low voltages ($8.2 < V < 8.5$ V, Fig. 3c), the solitons move directionally parallel to $\bar{n}_0$ in the $xy$ plane. Notably, they randomly move to the left or right because of the previously discussed dielectric oscillation. As plotted in Fig. 3i, at $V > 8.5$ V, the directional angle of motion with respect to $\bar{n}_0$ continuously varies as the voltage is increased (Figs. 3d–h), corresponding to the increasingly asymmetric deformation of the bug-eye-like pattern (Fig. 3b). High-speed camera observations reveal that this movement is driven by the oscillation of the director field between distorted versions of two orientational states, as shown in Fig. 1h. The directional motion ends at about 10 V, after which the soliton motion becomes somewhat chaotic, leading to a fractalised proliferation process that will be discussed later. The average velocity of the motion of the solitons is measured through a trajectory tracking as shown in Figs. 3c-d and plotted as a function of voltage and frequency in Figs. 3j, k. Although the velocity is nearly constant with respect to the voltage because the background flow field is mostly unaffected in the limited voltage range, a significant frequency dependence is observed. The solitons are resonantly active at about 20 Hz, as shown in Fig. 3k. This frequency corresponds to a critical frequency that separates the conductive and dielectric regimes[32], as described by the competition between the relaxation process of mobile charge in space and the dielectric relaxation of the director, $\sigma_{//}/\varepsilon_0\varepsilon_{//}$. Given $\sigma_{//} = 10^{-9}$ 1/Ω·m and $\varepsilon_{//} = 4$, as measured experimentally, we obtain the critical frequency of 28 Hz by calculation, consistent with the experimentally observed value.

The dynamics of collision between solitons and their reflection by obstacles can be understood in terms of their pseudoparticle nature. Figs. 4a–c show that their head-on collisional behaviour varies depending on the degree of offset, $\Delta$, between soliton cores. When $R < \Delta < 2R$, interference occurs between the peripheries of soliton pairs with the same orientation. Consequently, the solitons repel each other at a small post-collision angle; at $0 \leqq \Delta < R$, elastic deformation becomes more significant because of the increased mismatch of the director field between solitons, leading to vertical repulsion between the solitons. Following this collision, the solitons separate such that elastic deformation is avoided and then continue to move along their original directions of motion. Interestingly, the solitons are elastically reflected at an air–nematic interface in the manner of real particles colliding with a wall (Fig. 4d).

**Proliferation regime.** Figs. 5a and b show the fractalised proliferation trajectory of destabilised solitons for a 10-V field at 40 Hz. Soliton fractalisation produces clones for movement perpendicular to $\overline{\boldsymbol{n}}_0$. Upon fractalisation, the solitons bifurcate and move obliquely for a short distance. The fractalisation process corresponds to the continuous growth in soliton size. Considering that an optimum size exists for the solitons at a fixed film thickness (Supplementary Fig. S7), the fractalisation process results in the continuous accumulation of excess elastic energy. The solitons fractalise when the energy penalty from elastic deformation cannot be compensated by the effects of surface anchoring and dielectric interaction. This is illustrated in Fig. 5c, which shows the soliton diameter, $2R$, as a function of time; fractalisation occurs when the soliton size exceeds a specific value. Notably, the fractalisation simultaneously occurs for all solitons (Figs. 5a and b). This implies that a synchronised growth mechanism is involved for the solitons, resulting in a solitonic trajectory with a topological Hausdorff fractal dimension of $D = 1$.

**Discussion**

**Soliton kinetics.** We first discuss the coupling of the far-field background flow field with elastic deformation in the Soliton-state that drives the kinetics of the solitons. In the Soliton-state, convective flows with waves obliquely traveling to both the left and right relative to $\bar{\bm{n}}_0$ simultaneously appear, as confirmed by the observation of roll patterns coexisting with solitons in the ⊥-state. The localised dynamics are assumed to consist of two oblique waves traveling at an angle $\theta$ with respect to $\bar{\bm{n}}_0$, as introduced elsewhere[33-35]. As a first-order approximation, the amplitude expansion of the flow field is $w = [A(r,t)e^{ik_A \cdot r} + B(r,t)e^{-ik_B \cdot r}]e^{(ik_C \cdot r - \omega t)} \sin z$, where $A$ and $B$ are the respective spatiotemporal variable amplitudes of the two assumed traveling waves; $k_A$, $k_B$, and $k_C$ are the wave numbers; $r$ is the spatial coordinate; $t$ is time; and $\omega$ is the angular frequency. For small amplitudes $A(= A_0 e^{-i(Qx + Py - \Omega t)})$ and $B(= B_0 e^{-i(Qx - Py - \Omega t)})$, the description of the flow waves in the Ginzburg–Landau framework is valid, yielding $\frac{\partial}{\partial t} A = \frac{\partial}{\partial A}[-\int dV(\bm{u}_A \cdot \nabla A) + (\mu - c|A|^2 - g|B|^2)A^2 + (\nabla A)^2]$ and $\frac{\partial}{\partial t} B = \frac{\partial}{\partial B}[-\int dV(\bm{u}_B \cdot \nabla B) + (\mu - c|B|^2 - g|A|^2)B^2 + (\nabla B)^2]$, where $\bm{u}_{A,B}$ are the group velocities of the waves, $\mu$ is the bifurcation parameter, and $c$ and $g$ are coupling coefficients. Forming a symmetric superposition mode of the two waves in the *xy* plane with a finite $\Omega$ proportional to $|A_0|^2$ produces a two-dimensional localised traveling rectangle solution corresponding to a solitonic state. If only the flow field is under action, two types of solitonic motion are possible: a directional motion parallel to $\bar{\bm{n}}_0$ and a two-dimensional random walk within the *xy* plane. However, in our results, variable angles of solitonic motion with respect to $\bar{\bm{n}}_0$ are achieved. This motivates us to consider the elastic deformation of the bulk directors by introducing a spatially modulated imbalance of amplitudes between waves $A$ and $B$. Although convective flow is considered in Ref.

[33], our solitons have splay deformation in the *xy* plane and twist-bend deformation in the *xz* plane (Fig. 2i) in addition to a convective flow. Owing to the flow-alignment property of the director, its orientation tends to be parallel to the flow field. Since the orientation of the director oscillates with time, the local flow also changes its direction accordingly through a self-regulating procedure (Supplementary Fig. S9). Such processes can trigger the local motion of topological defects[36], thereby providing an additional traveling mode out of the *xz* plane. Indeed, the electrohydrodynamic convection flow can be modulated by twist deformation to produce a spiral flow[37]. Because the elastic deformation in the *xy* plane weakens in the transition from the ∥-state through the Soliton-state to the ⊥-roll state as the frequency is increased or the voltage is decreased, the inclination of the oblique solitonic movement decreases with respect to $\overline{\boldsymbol{n}}_0$; with neither flow nor elastic deformation in the *xy* plane, no two-dimensional motion is observed in the studied parameters (Figs. 2a–e). Thus, the local coupling between the elastic deformation and the flow field in an anisotropic medium permits effective dictation of the solitonic motion directionality, which cannot be achieved through a flow field alone.

**Pseudoparticle soliton properties.** Herein, we discuss the pseudoparticle properties of solitons in anisotropic flows. For characterisation, we analyse a radial distribution function of packed solitons, $g(r_{cc})$ (Supplementary Fig. S9) and the thermal-energy-activated solitonic Brownian diffusion. Fig. 6a shows the pairwise potential $V(r_{cc})$ as a function of the center-to-center distance between solitons, $r_{cc}$. At the dilute limit, $V(r_{cc})$ is directly calculated as $\frac{V(r_{cc})}{k_B T} = -\ln[g(r_{cc})]$. The first potential closely corresponds to the mean nearest-neighbor distance between solitons, $r_{cc,o} = 15.2$ µm. The depth of the potential well slightly exceeds the thermal energy, $k_B T$, which is consistent with the observed minimal soliton fluctuation beyond $r_{cc,o}$. The solitonic Brownian diffusion process is calculated as the two-dimensional mean square displacement of the solitons over time $t$ (Fig. 6b). The effective diffusion constant of the solitons is 1.4 µm²/s, corresponding to a mean solitonic fluctuation velocity of about 1 µm/s. The effective mass of the solitons, $m_{eff}$, is expressed as $U = 1/2 m_{eff} v^2$, where $U$ is the kinetic energy and $v$ the mean velocity of solitonic fluctuation. By balancing the thermal energy, $k_B T$, and the kinetic energy of the solitons, $m_{eff}$ is estimated to be 1 pg. In contract, the real mass of molecules constituting each soliton is estimated to be $\rho \pi R^2 h \approx 1000$ pg, where $\rho$ and $h$ are the material density and nematic film thickness, respectively. The effective mass of the soliton is significantly smaller than the real mass of the materials constituting the solitons. Notably, $m_{eff}$ is independent of the film thickness irrespective of the change in the real mass (Supplementary Fig. S12). Because the elastic energy density decreases as the film thickness increases, the total elastic energy does not vary. This means that $m_{eff}$ reflects the elastic energy stored by the elastic deformation in the director field.

**Summary**

We developed and demonstrated a strategy for manipulating the complex spatiotemporal behaviour of solitons by fine-tuning several parameters upon which the coherent interactions between the traveling flow and elastic deformation are highly dependent. This system provides a model for observing and controlling soliton dynamics and suggests that solitary behaviours reflect a more universal nature applicable to other soft-matter systems. Further work is required to determine whether additional parameters such as chirality and local ordering drive new solitonic behavioural features.

**Methods**

**Sample preparation.** The frustrative nematic mixtures were formed by mixing E7 (a mixture of cyanobiphenyl and cyanoterphenyl components, Wako, Aldrich, and made in house) with 4′-butyl-4-heptyl-bicyclohexyl-4-carbonitrile (CCN47, Nematel GmbH) by weight. These molecules form a uniaxial nematic phase in which the average molecular orientation is characterised by a vectorial director, ***n***. Because of the macroscopic and nonpolar nature of the nematic phase, the head and tail of ***n*** are equivalent. The liquid crystals are uniformly aligned between two glass plates with a controlled thickness in the range of 2–50 µm. In the manuscript, if not specified, we show solitonic behaviours in a 5.2-µm-thick film. Although E7 has positive anisotropies $(\Delta\varepsilon\Delta\sigma) = (++)$, those of CCN47 are negative $(\Delta\varepsilon\Delta\sigma) = (--)$. Tetrabutylammonium benzoate (TBABE) was added to the E7-CCN47 mixture to tune its net strength of conductivity. Films comprising these mixtures were sandwiched between pairs of substrates coated with rubbed polyimide on indium-tin-oxide electrode layers, with the film thickness adjusted by adding silica particles to obtain a 2–12-µm-thick film (micromer, Micromod) or by adding polyester film spacers to obtain a 20–50-µm-thick film (Mylar, Dupont). In many other liquid crystalline systems, including Schiff bases and cyanobiphenyl mesogens, solitons were also found.

**Conductivity measurement.** The electrical and dielectric responses in the liquid crystal mixtures were characterised via dielectric spectroscopy (Solartron Analytical, impedance/gain-phase analyser 1260A with dielectric interface system 1296A). An AC electric field at 0.05 $V_{rms}$ sweeping at a frequency of $10^{-2}$–$10^4$ Hz was applied to the films, and the resulting impedance response was recorded and analysed (Supplementary Fig. S13).

**Director mapping.** The orientation of the directors was determined optically using polarising microscopy. The highest birefringence of the mixtures was sufficiently low (0.03 < $\Delta n = n_e - n_o$ < 0.08, where $n_e$ and $n_o$ are the extraordinary and ordinary refractive indices, respectively) to carry out this task. On the basis of the measured transmittance of light through the films, we calculated the orientation of the directors. Our method was the same as that presented in Ref. [21] . We also conducted fluorescence confocal polarising microscopy (TCS SP8 STED, Leica) to probe the three-dimensional distribution of the soliton directors. Although volume scanning was performed during soliton movement, owing to their fast fluctuation that was synchronised with the electric field, the solitons were not clear enough to allow their analysis. Instead, we measured the emission intensities at single planes on the surfaces of the slab and at the middle of the cell. The intensity on the surfaces was nearly constant during the application of the electric field, confirming a strong surface anchoring effect. The signal was maximum at the middle the slab, corresponding to elastic deformation of nearly zero on the surface that reached a maximum at the middle of the slab.

**Characterisation of flow field.** The particle tracking method was used to visualise the effective flow in the mixtures in the presence of an electric field. In this process, a small number of 2-µm particles were dispersed into the nematic mixtures, and their motions were tracked by both bright-field and polarising microscopy.

**Time-dependent trajectory analysis.** The recorded dynamics of the solitons were analysed using custom-made ImageJ macros. For the precise image analysis needed to extract sharp shapes and calculate the centers of gravity of the solitons, a dark but slightly structured time-dependent background was averaged over time.

**Acknowledgments**

The authors thank H. Orihara (Hokkaido University), A. Buka (Hungarian Academy of Science), P. Salamon (Hungarian Academy of Science), H. Yoshida (Osaka University), I. I. Smalyukh (University of Colorado Boulder), and J. Yoshioka (Ritsumeikan University, ex. RIKEN CEMS) for their valuable input. We would like to thank Editage (www.editage.jp) for English language editing.

**Author contributions**

S.A. and F.A. designed and directed the research, discussed the mechanisms, and wrote the manuscript. S.A. performed all the experiments and analyses.

**Additional information**

Supplementary information is available in the online version of this paper. Reprints and permissions information is available online at www.nature.com/reprints. Correspondence and requests for materials should be addressed to S.A. or F.A.

**Competing financial interests**

The authors declare no competing financial interests.

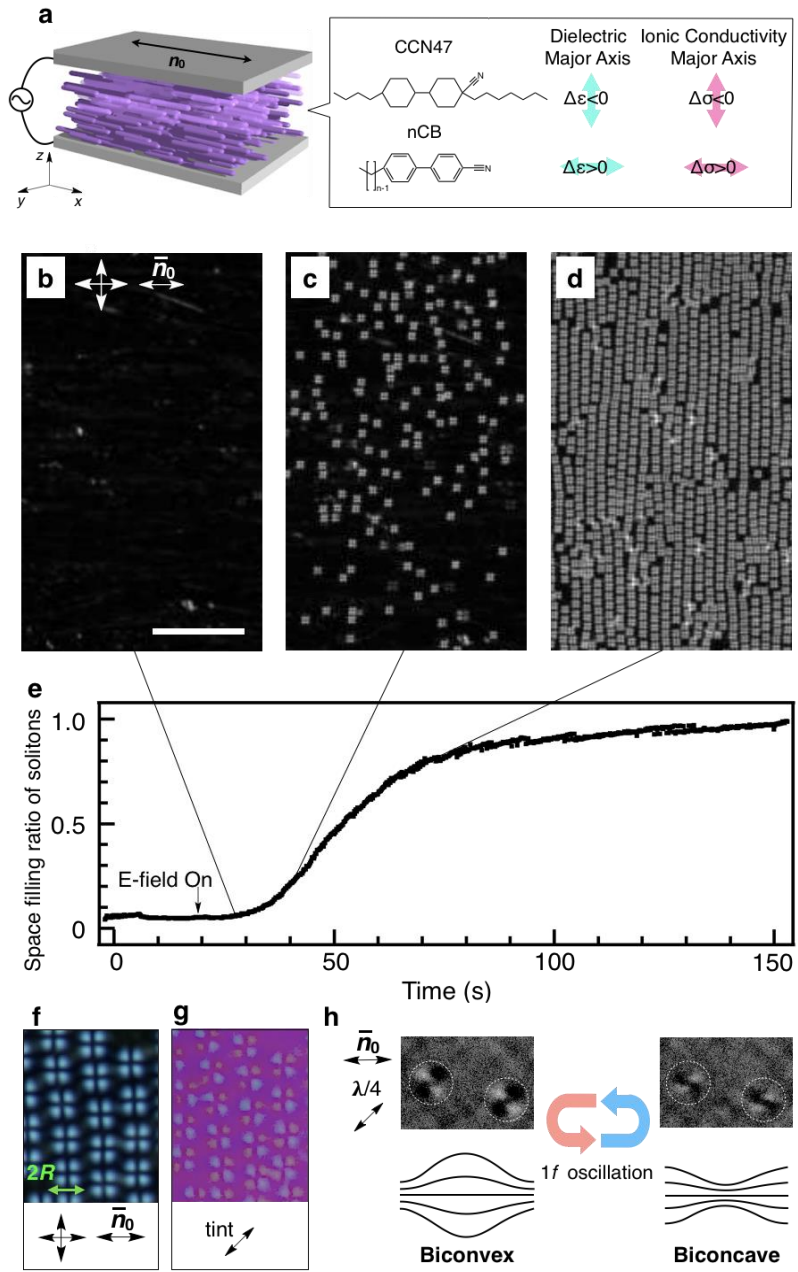

**Figure 1. Creation of ordered solitons.** (a) Slab geometry and chemical structures of the main materials—cyanobiphenyl (CB) and 4'-butyl-4-heptyl-bicyclohexyl-4-carbonitrile (CCN47). (b–d) Nucleation and growth processes of electrically pumped solitons visualised using polarising microscopy with crossed polarisers for specific times indicated by the time evolution of the soliton space-filling ratio in (e). Scale bar, 100 µm. (e) Time evolution of the space-filling ratio of solitons. (f) Polarising microscopy images of packed solitons. The soliton radius, $R$, is defined as the radius of the primary area undergoing elastic deformation.

The space-filling ratio $X$ is defined as the ratio of the total area of $N$ solitons to the sample area, $A$. (g) Counterpart to (f) taken with a tint plate. (h) Monochromatic textures with a $\lambda/4$ plate and schematics of the alternating director field between two orientational states, taken with a high-frame-rate camera.

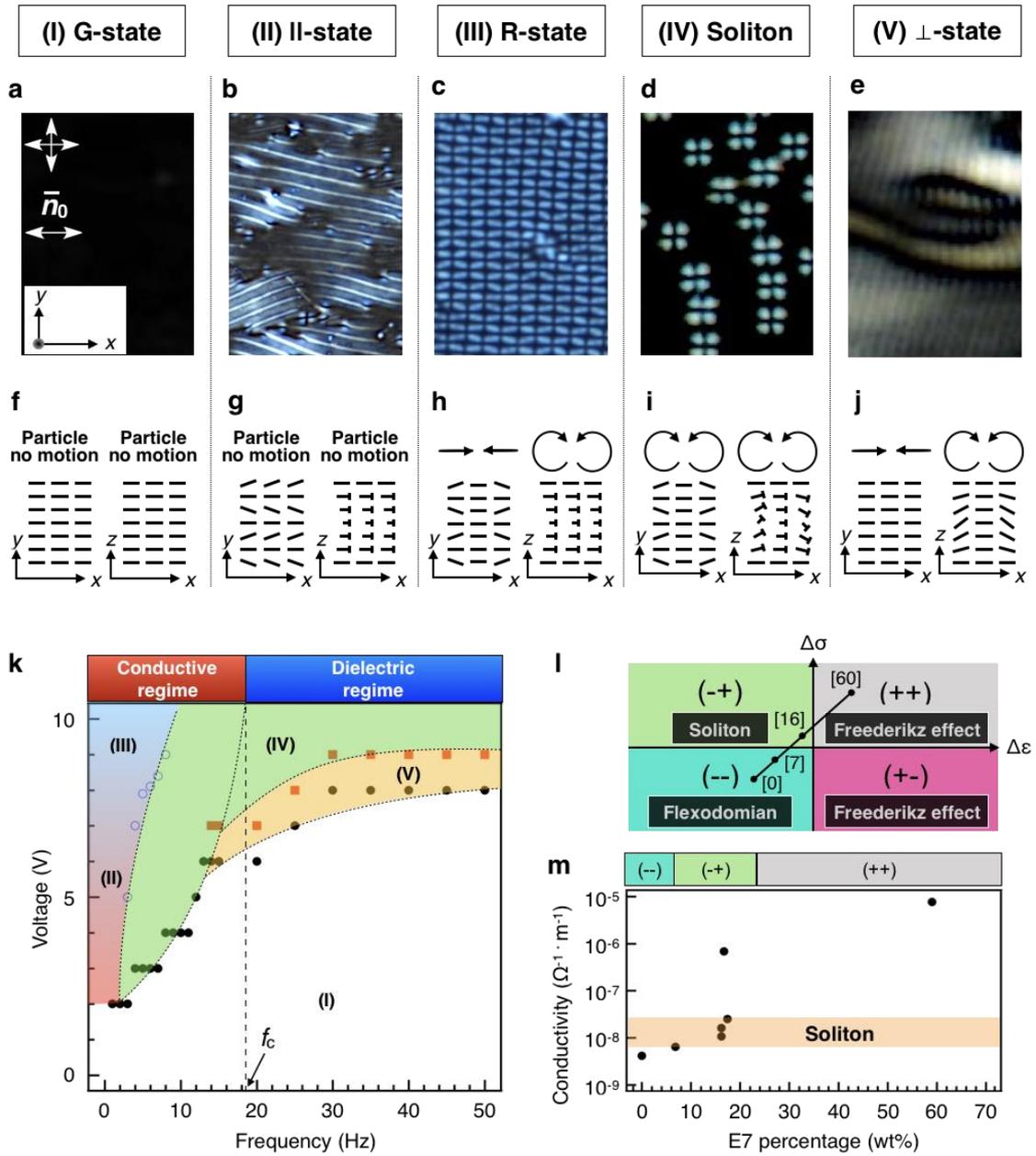

**Figure 2. Conditions for generating solitons.** (a–e) Polarising microscope textures of the topological states in the presence of an electric field. (f–j) Schematics of the director field by nail notation and motion trajectories of particle motion. The director field in the *xy* plane is on the middle plane of the cell. (k) State diagram as a function of the voltage and frequency. (l) Primary phenomena occurring because of the changes in the signs of the anisotropies of dielectricity ($\Delta\varepsilon$) and conductivity ($\Delta\sigma$). The solid line indicates a route for tuning the combination of anisotropies by mixing E7 at different weight percentages, which are shown in square brackets, with CCN47. (m) Change in the specimen conductivity obtained by changing the concentration of E7 in CCN47.

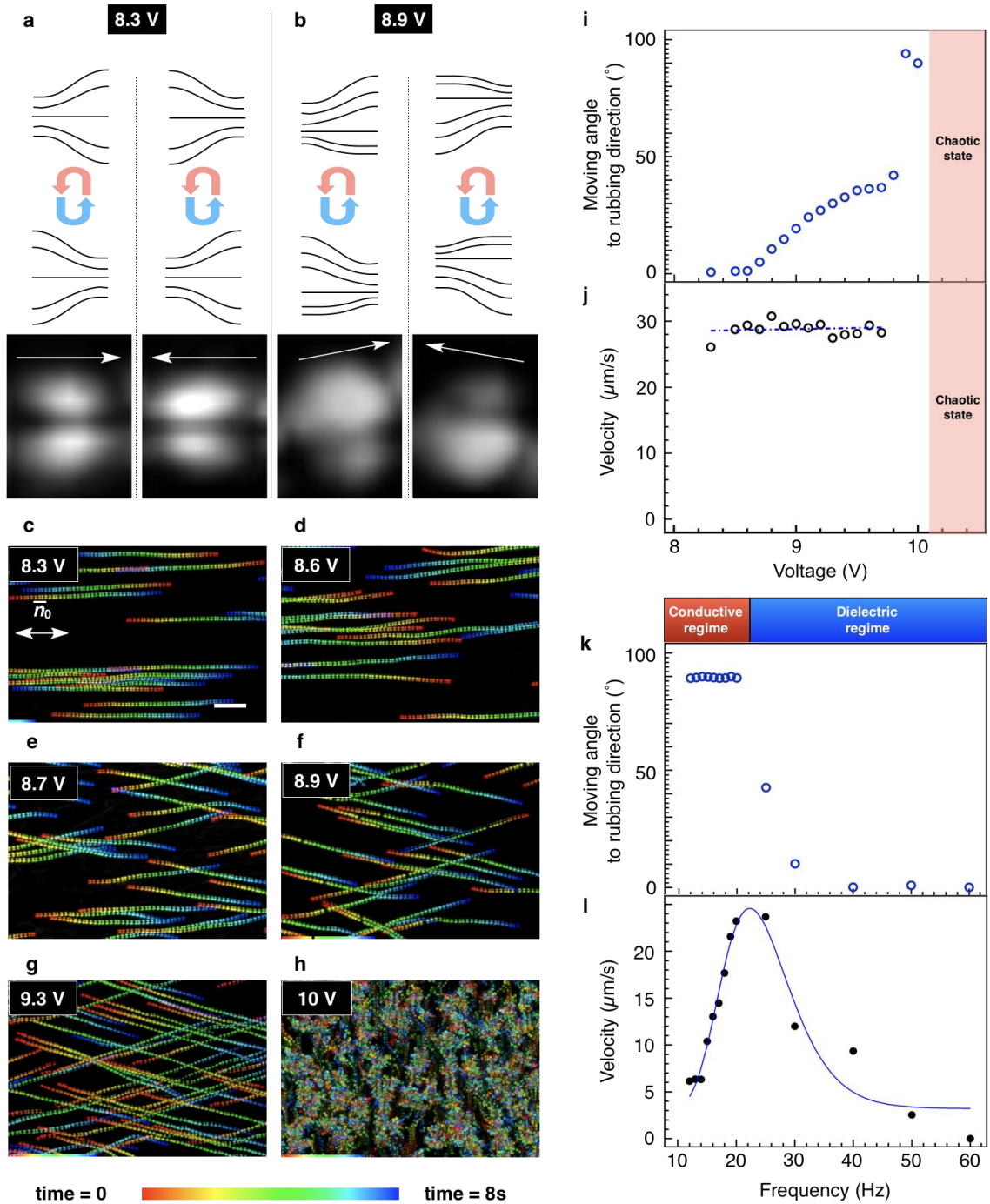

**Figure 3. Time-dependent dynamics versus electric-field parameters.** (a, b) Director distortion of the kinetic solitons at 8.3 V and 8.9 V. Schematics of the alternating distorted convex and distorted concave director fields are presented with magnified patterns of them. (c–h) Typical time-dependent trajectories of solitons during different motions at various voltages at 20 Hz, stacked over 8 s. Scale bar = 50 µm. (i) Angles of soliton motion with respect to $\overline{n}_0$ at 20 Hz. (j) Velocity of the solitons at 20 Hz plotted as a function of the voltage. (k) Angles of soliton motion with respect to $\overline{\boldsymbol{n}}_{\boldsymbol{0}}$ at 10 V Hz as a function of the frequency. (l) Velocity of solitons at 10 V as a function of the frequency.

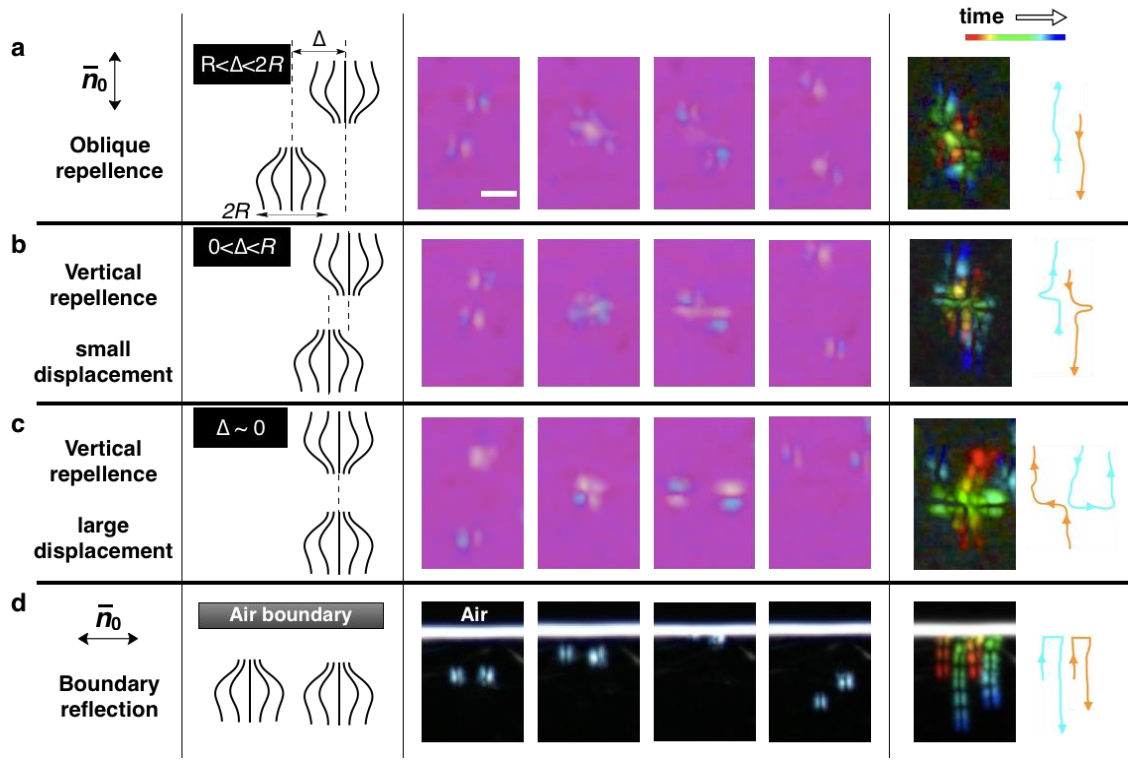

**Figure 4. Elastic collisional trajectory tracking.** (a–c) Snapshots of collisions between solitons (with a tint-plate) and (d) reflection of solitons by an air boundary (with crossed polarisers) with schematics of solitonic topologies. In the second column from the left, the schematics of distorted biconvex director fields before collisions are drawn. Instantaneously, the director field alternates between distorted versions of two orientational state, as shown in Fig. 3a. Dashed lines in the schematics are used to clarify the positional offset, Δ. The overall trajectories are shown with colour-code traces and arrows in the right-most column. Scale bar, 10 µm.

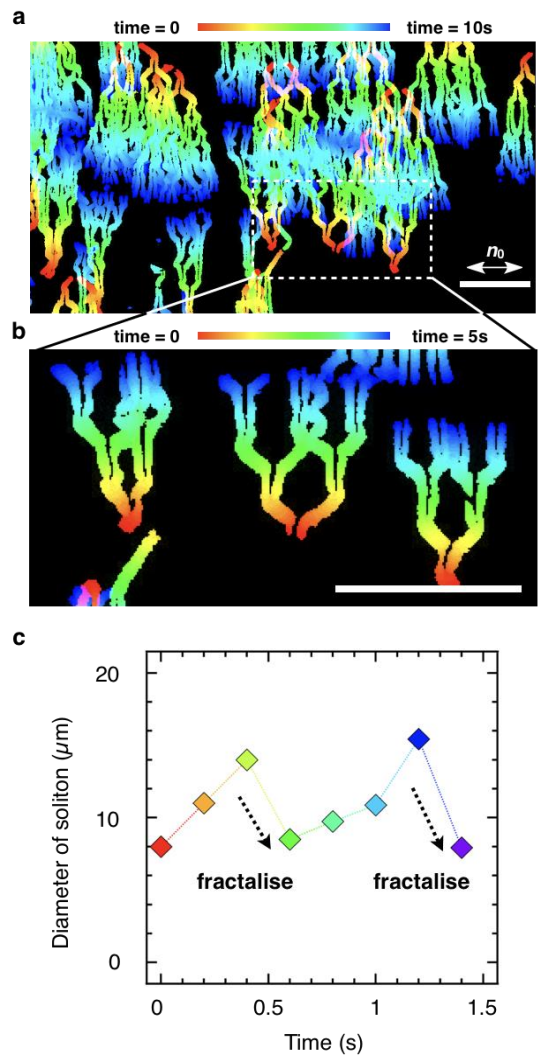

**Figure 5. Fractal cloning of solitons.** (a) Time-dependent trajectories of solitons during fractal cloning, stacked over (a) 10 s and (b) 5 s. Scale bars, 100 µm. (c) Fractalisation occurring when the soliton size is nearly doubled.

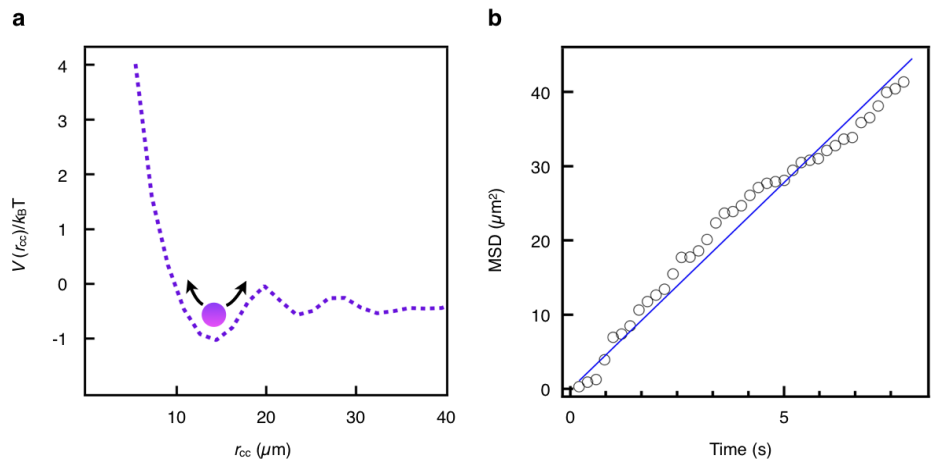

**Figure 6. Pseudoparticle properties of solitons.** (a) Pairwise soliton potential. (b) MSD of solitons as a function of time.

# Supplementary Information

Satoshi Aya*,# and Fumito Araoka*

## Table of Contents

---

\* To whom correspondence should be addressed.
E-mail: satoshi.aya@riken.jp (S.A.); fumito.araoka@riken.jp (F.A.)

# Current contact address: South China University of Technology,
E-mail: satoshiaya@scut.edu.cn

**1. Ion localisation in solitons**

Ions play important roles in stabilising solitons. Here, we show an essential experimental result regarding ion localisation in solitons, i.e., the observation of the redistribution of ions. In the experiment, we first switch on an electric field to create solitons. At an arbitrary time, we remove the electric field, and the solitons instantly disappear. Then, we wait for *X* seconds for the next application of the electric field to create solitons. If a localisation process of ions is involved in soliton generation, the diffusion or redistribution of ions should occur after the electric field is removed. We observe this by tracking the location at which the solitons are regenerated while tuning the waiting time *X*. As a result, we find that solitons are regenerated at the same location as before the electric field is removed if *X* is less than about 20 s (Fig. S1); otherwise, they are randomly regenerated in space. This means that some ions are indeed accumulated at the center of the solitons upon their generation. Once the diffusion of ions is complete after the electric field is removed, new solitons are randomly created in space at the second time of generation because there are new seeds for the solitons due to the localisation of ions. Taking a typical diffusion coefficient of the ions in liquid crystals of $2\times10^{-9}$ m/s$^2$, it is calculated that the delocalisation of the ions takes about 20 s, which is consistent with the observed time scale (Fig. S2).

**2. Analysis of the soliton growth process**

A Kolmogorov–Johnson–Mehl–Avrami (KJMA) analysis of the structure-forming events provides information on how homogenously these events unfold and how their nucleation and growth processes proceed[S1–S4]. To estimate the occurrence frequencies of the electrically pumped solitons, we plotted the time evolution of the soliton volume ratio during their creation using the Avrami expression, $ln[ln\,1/\{1-\Phi\}] = m\,ln\,t + ln\,K$, where $X$, $t$, $K$, and $m$ are the soliton volume fraction, time, temperature-dependent Avrami coefficient, and Avrami exponent, respectively (Fig. S3). We found that the overall creation process presents a good fit with the Avrami scale with an Avrami exponent of 2.9. Considering the random incidence of each soliton within space, as revealed by polarising microscopy, the creation of solitons can be seen to be mediated by the homogeneous nucleation of soliton seeds, which is initiated by a localisation of ions, as discussed later, and the subsequent two-dimensional spreading of their network.

Now, we turn to consider why the KJMA model fits our data well. When creating solitons, switching on an electric field initiates the localisation of ions, and the localised ions becomes seeds that have potential to trigger the creation of the solitons. Let us assume the number of seeds at time 0 is $N_0$ and these can be activated to create solitons through a first-order-like reaction process. Then, the number inactivated seeds at time $t$ can be written as $N_0 exp(-at)$, and the activation rate of the seeds to be nucleus of the solitons per unit time $aN_0 exp(-at)$. As a result, the emerging number of solitons is proportional to $N_0 - aN_0 exp(-at) \propto 1 - exp(-kt^n)$, which is in a same expression to that of the KJMA model and validates the usage of the model to evaluate the growth process of the solitons.

## 3. Packing of solitons

We analysed images of packed solitons to gain a better understanding of their packing states. Figure S4a shows a swarm of solitons packed in a centered rectangular manner. A fast Fourier transform of Fig. S3a is shown in Fig. S3b, and the corresponding direct observation via the conoscopy of light diffraction from the packing state of the solitons is shown in Fig. S3c. It is obvious that the fast Fourier and conoscopy results are consistent.

**4. Structure of solitons**

As stated in the main manuscript, the solitonic structure is dynamic. When we observe the structure of the solitons under a POM by a conventional low-speed camera, which has a capturing rate lower than the oscillating frequency of the dynamic solitons, the solitons appear as a time-averaged structure. In this case, they look like bearing topological defects near the their center. In order to clarify the time sequence of the structure, we perform high-frame-rate polarising microscopy with a $\lambda/4$ plate at 100 fps. In Fig. S5, the time evolutions of the transmittance in two domains of the observed textures are plotted as a function of time along with a schematic director field and the textures at specified times. $T$ is the periodicity of the applied electric field. Clearly, it is shown that the transmittance of the biconvex structure is stronger than that of the biconcave one. This suggests that the maximum deviation angle of the director in the biconvex structure is larger than that in the biconcave one. Another key feature is that the inclination angle of the brushes in the textures between the biconvex and biconcave structures is very different, indicated by the white dotted lines. Lastly, both the switching times from the biconvex to the biconcave and from the biconcave to the biconvex are of the order of several millisecond, and the switching time from the biconvex to the biconcave is shorter. These suggest the structural oscillation is elasticity-driven. All of these evidences confirm that the biconvex and biconcave structures are not a simple symmetric inversion of the director field, but the director field is more distorted in the biconvex structure.

## 5. Conditions of soliton creation: comparison between the current solitonic systems and conventional convective systems.

The state diagram (Fig. 2a-e, Figs. S6,S7) infers an important connection between the current solitonic systems and conventional convective systems. We compare the known state diagram in a review by Eber et al.[S5] (Fig. S6, Left) with our state diagram (Fig. S6, Right). It turns out that the G-, ∥-, R- and ⊥-states appear in similar ranges with the ones in conventional convective systems. The significant difference is that the Soliton state replaces the travelling waves and dielectric rolls regimes, and suppresses the chaotic regime.

## 6. Kinetics of solitons

In order to show the details of the kinetics of solitons, we show a state diagram in Fig. S7, focusing on the soliton state. The dynamic solitons just fluctuate around their equilibrium position at low frequencies (orange regime). At low frequencies and high voltages (Purple regime), the solitons swim parallel to the direction of the background director. At higher frequencies and high voltages (Blue regime), the solitons swim perpendicular to the direction of the background director. Since this regime is right next to the electrohydrodynamic regime ((V) ⊥-state) where the direction of background flow is parallel to the direction of the background director, the swimming direction is dominantly determined by the flow as we stated in the Discussion part. In the narrow shaded regime between the blue and orange regimes, the solitons change their swimming direction between parallel and perpendicular to the direction of the background director, as shown in Figs. 3i-l. In this regime, the background flow becomes slightly oblique, suggesting an additional travelling mode out of the *xz* plane becomes stronger.

**7. Go-and-stop motion of swimming solitons**

We observe the trajectory of the solitons upon their swimming, taken by a high-speed camera. This allows us to analyse the accurate dynamics of the solitons. Fig. S8a demonstrates an increment of *x*-coordinate of a soliton as a function of time when the soliton is moving along *x* direction. It is clear that *x*-coordinate of the solitons averagely-increases, and oscillates. Fig. S8b shows a power spectrum of Fig. S8a, revealing the rate of the go-and-stop motion is about 16.5 Hz, consistent with the frequency of the electric field, 16 Hz. This fact suggests that the go-and-stop motion is caused by the switching between biconvex and biconcave structures upon solitonic swimming.

**8. Size of solitons**

The size of solitons is determined by the balance between the elastic energy and the surface anchoring. As discussed in the manuscript, the solitons carry out-of-plane deformation along the *z* axis. Therefore, it is expected that the size of the solitons is dependent on the sample thickness; that is, a thicker sample results in larger solitons. In Fig. S9, the linear dependence of the size of the solitons on the sample thickness is shown.

**9. Feedback processes of the director and flow field through a flow-alignment process**

The orientation of the director tends to be parallel to the flow field as a result of the flow-alignment property. Since the flow field deviates from the initial state after the reorientation of the director, the orientations of both the flow field and director oscillate as a result of a self-regulating procedure (Supplementary Fig. S10).

**10. Discussion of soliton stabilisation**

Herein, we discuss why dynamic solitons have been so difficult to observe, despite the lengthy history of research on the electrohydrodynamics of liquid crystals, and, correspondingly, why our system is a special case. We again emphasise that the key property of the soliton system examined in this study is the frustration of dielectricity and conductivity. As shown in Fig. 2l, the Soliton-state appears only for $(-\ +)$ mixtures at moderately low conductivities in the range of $8 \times 10^{-9} < \sigma < 4 \times 10^{-8}\ \Omega^{-1} \cdot m^{-1}$, which is two or three orders of magnitude lower than that of typical materials used in electrohydrodynamic studies but similar to previously reported values[S6,S7]. The importance of this middle conductivity region can be understood by considering the coupling between the space charge $Q$ and the electric field $E$. In the lower-conductivity region ($\propto Q$), the number of free ions is spatially limited; as a result, the viscous torque proportional to the time-averaged quantity $<QE>_T$[S8] must destabilise the ground planar state, but is too small to overcome the total stabilising force produced by surface anchoring and dielectric alignment in the framework of the Carr–Helfrich electrohydrodynamic effect. In contrast, in the high-conductivity regime in which $Q$ is high and uniformly distributed in space, a viscous torque dominates and produces a turbulent flow. In the intermediate region, there is a sufficient number of ions for the Carr–Helfrich electrohydrodynamic effect to apply, but these can still be spatially localised. This consideration is well supported by the numerical calculation of electroconvective dynamics by Treiber and Kramer[S9]. Overall, the value of <$QE$>$_T$ effectively reflects the appearance of spatially localised waves corresponding to the Soliton-state.

**11. Radial distribution of solitons in the centered rectangular packing state**

The radial distribution function of the solitons in the centered rectangular lattice, $g(r_{cc})$, was calculated using imaging analysis. The radial distribution function is calculated as the ratio of the average soliton density, $\rho(r_{cc})$, at a distance of $(r_{cc})$ to the average density of solitons over the entire structure, $\rho$, as $g(r_{cc}) = 2\rho(r_{cc})/\rho$ (Fig. S11).

**12. Relationship between the solitonic trajectories and the pair-wise potential curve**

In Fig. 4, it is shown that the solitonic dynamic behaviour upon a collision is changed depending on the positional offset, Δ, of the solitons. Especially, when the positional offset, Δ, of two solitons in a collision event is about zero, i.e., a head-on collision, the solitons repels and make a vertical repulsion by a displacement of about 15 μm. This displacement is consistent with the value of the mean nearest-neighbor distance between solitons that is calculated by the pairwise potential curve (Fig. 6a). From this perspective, the collision is surely triggered by an elastic repulsion between the solitons.

**13. Characterisation of the electrical and dielectric properties**

The electric and dielectric responses of the liquid-crystal mixtures were characterised using dielectric spectroscopy (Solartron Analytical, impedance/gain-phase analyser 1260A with dielectric interface system 1296A). We focused on the low-frequency range (0.01–$10^4$ Hz), in which the main response originates from the charge polarisation of the mobile ions[S10]. The mixtures were confined between parallel glass plates with thin layers of indium tin oxide and were subjected to a 0.05-$V_{rms}$ AC electric field sweeping a frequency range of 0.01–$10^4$ Hz. Fig. S13 shows the dielectric spectra produced at various weight ratios of E7 to the host CCN47. Clearly, an increase in the ratio of E7 increases both the ionic relaxation frequency and dielectric strength as a result of the increased number of movable ions. Using the ionic characterisation method introduced by Sawada et al.[S10], we calculated the conductivities of the mixtures, as shown in Fig. 2m in the main manuscript.

**Supplementary Figures**

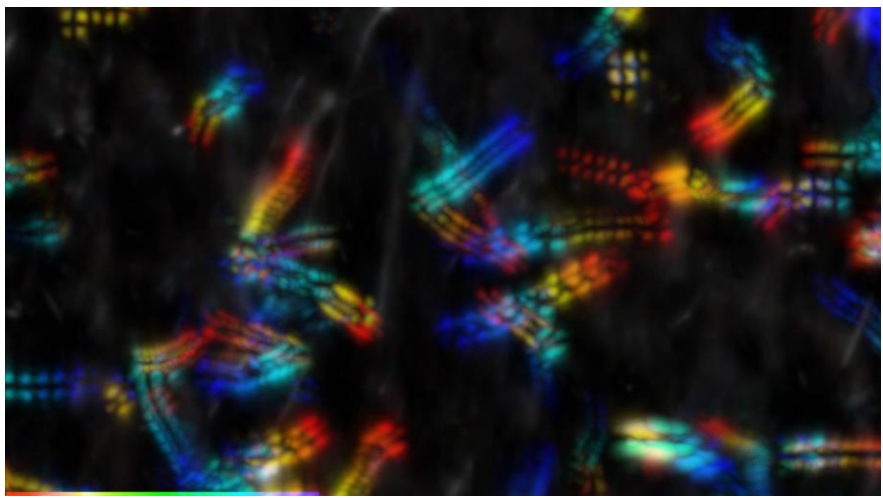

Figure S1. Trajectories of solitons. At the time (indicated by orange) where the first application of the electric field is removed, the solitons instantaneously disappear. Within 5 s (<X>), the electric field is applied a second time, and the solitons appear at same locations at which they disappeared when first application of the electric field was stopped.

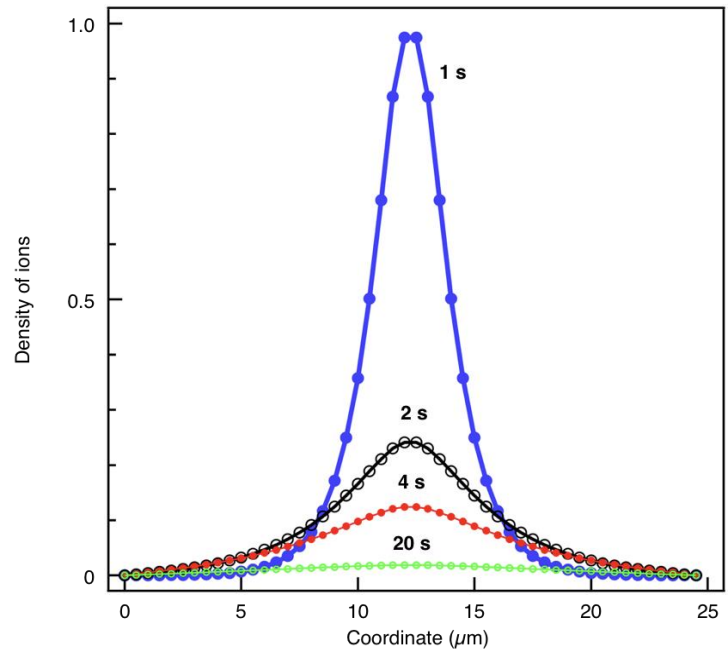

Figure S2. Simulation results for the diffusion of ions in space. At an initial state at $t = 0$ s, the ions are assumed to be localised at 12.5 µm upon the generation of solitons. The curves of the spatial distribution of the ions at 1, 2, 4, and 20 s are given in blue, black, red and green colours.

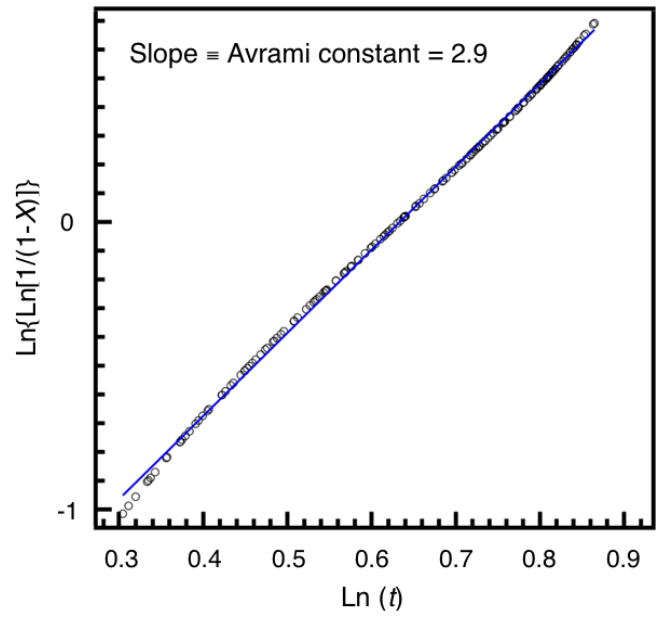

Figure S3. Avrami analysis of soliton creation with the time-dependent soliton volume ratio plotted on the Avrami scale.

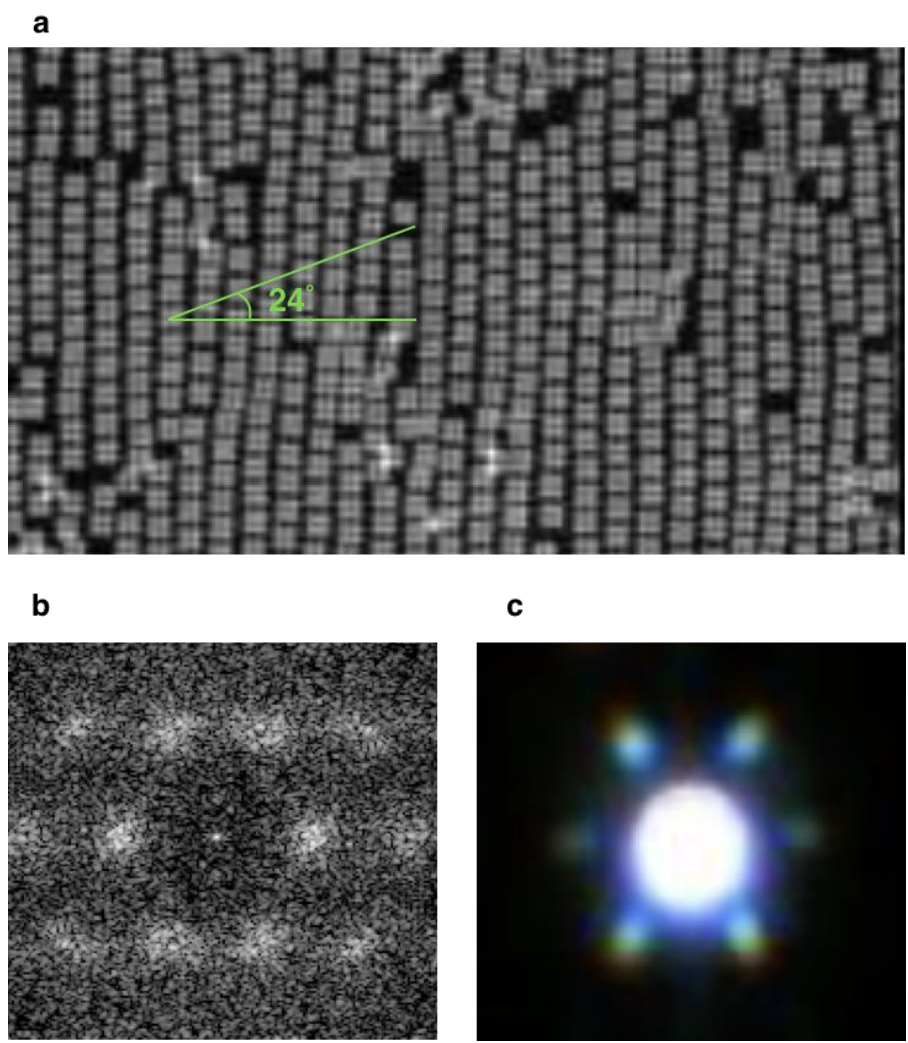

Figure S4. (a) Centered rectangularly packed solitons. (b) Fast Fourier transform of (a). (c) Conoscopy image of (a).

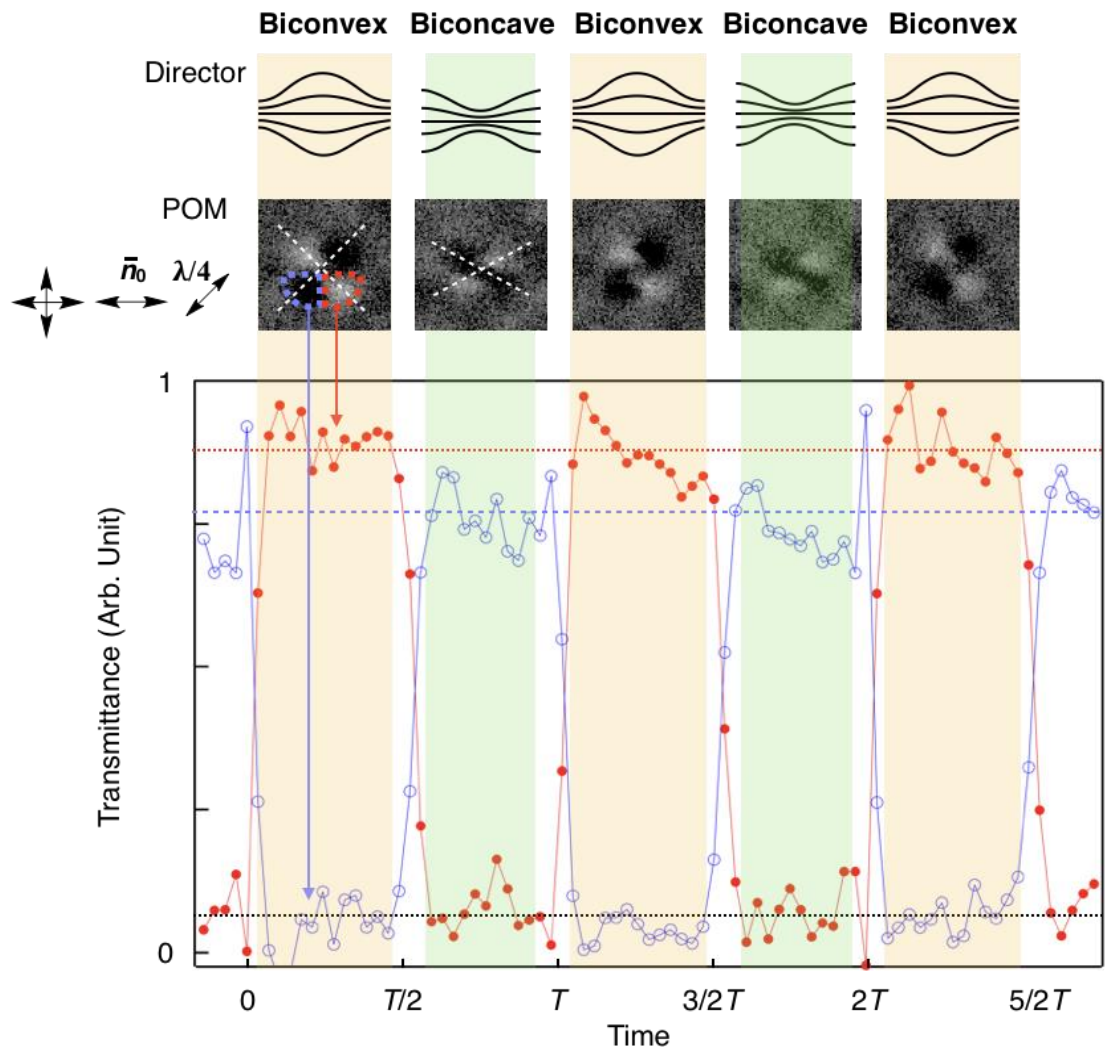

Figure S5. Time evolution of the structural variation in the solitons observed with high-frame-rate polarising microscopy with a $\lambda/4$ plate.

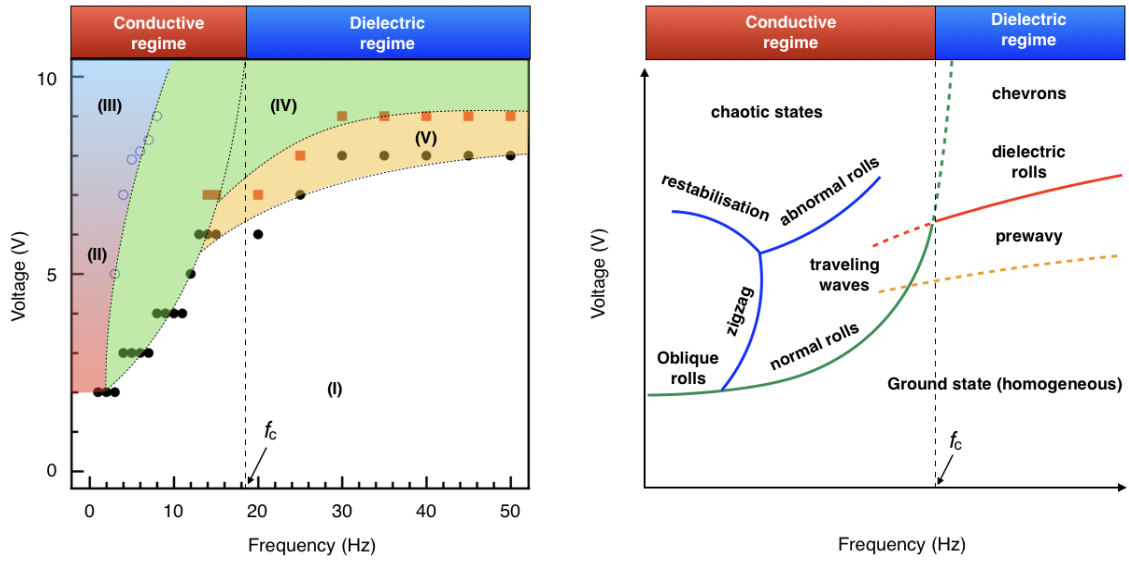

Figure S6. Comparison between the state diagrams of the current soliton system (Left) and conventional convective systems (Right, Reproduced from Éber et al.[s5]).

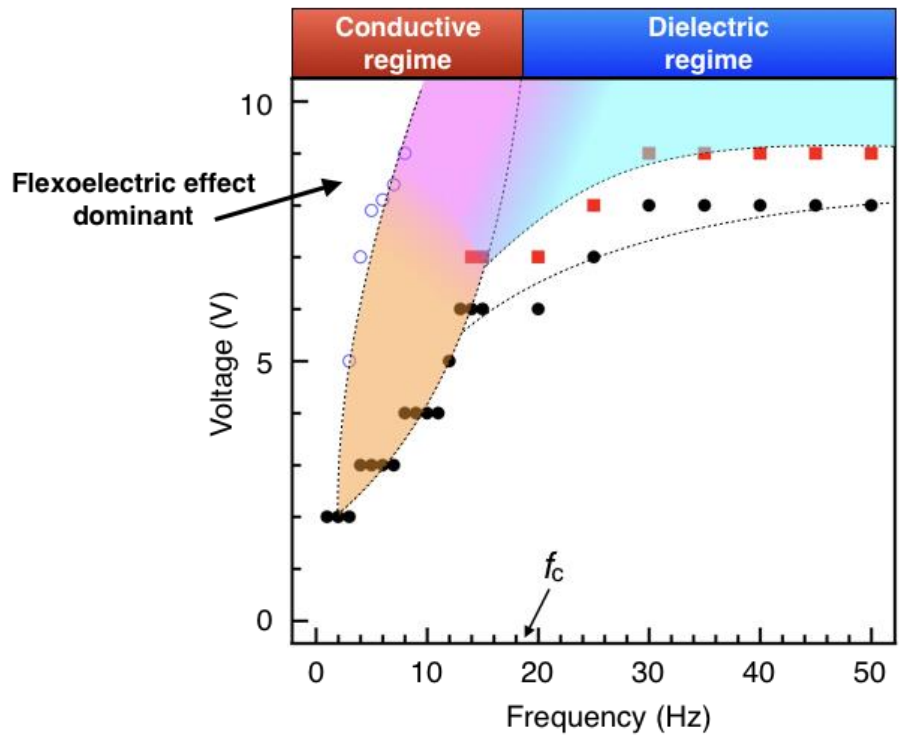

Figure S7. State diagram showing different regimes in the soliton state.

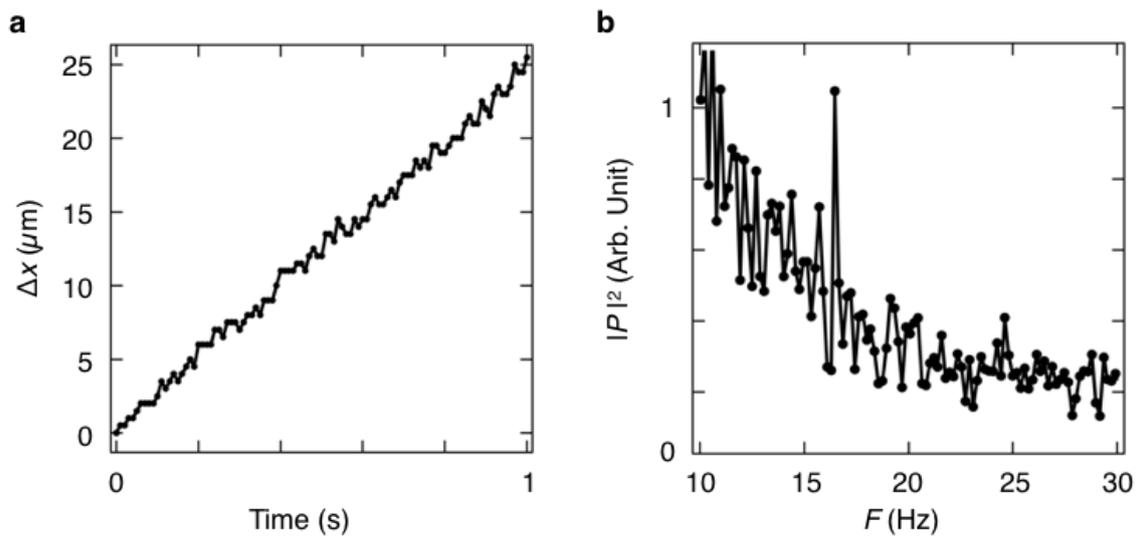

Figure S8. Analysis of the go-and-stop motion of the solitons at an electric field of 16 Hz.

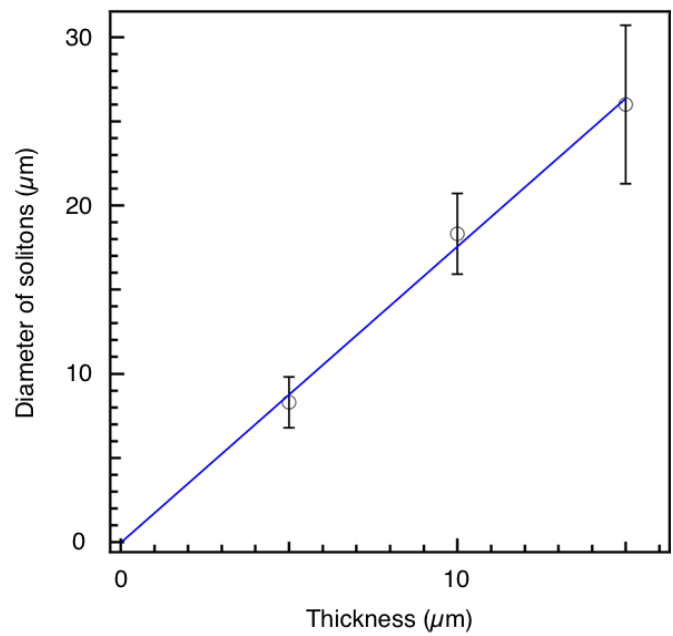

Figure S9. Thickness dependence of the diameter of the solitons.

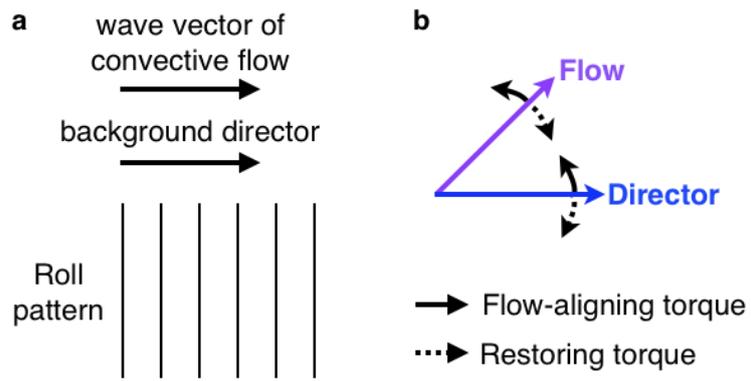

Figure S10. (a) Roll pattern in relation to the wave vector and the direction of director.

(b) Schematic of the feedback mechanism of oscillation of the director and flow field.

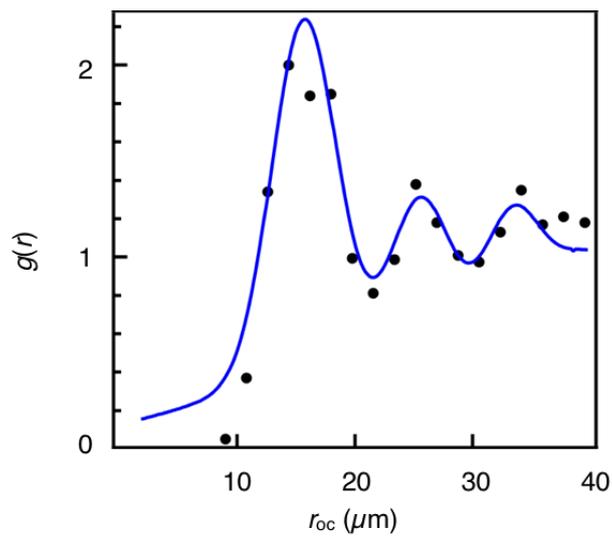

Figure S11. Radial distribution function of centered rectangularly packed solitons.

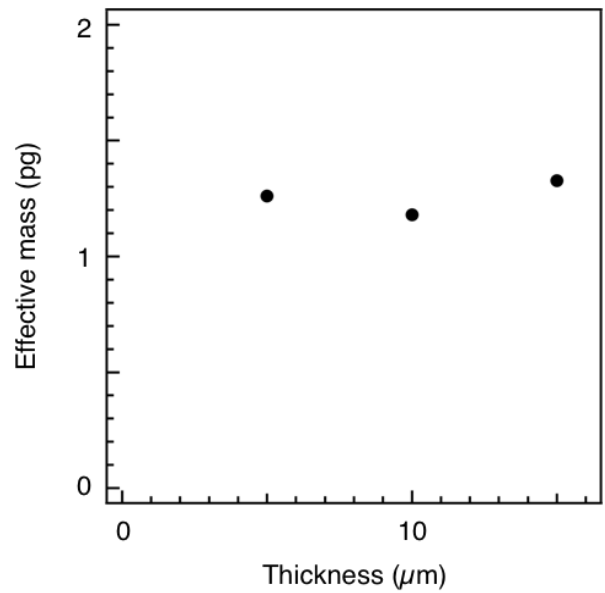

Figure S12. Thickness dependence of the effective mass of the solitons.

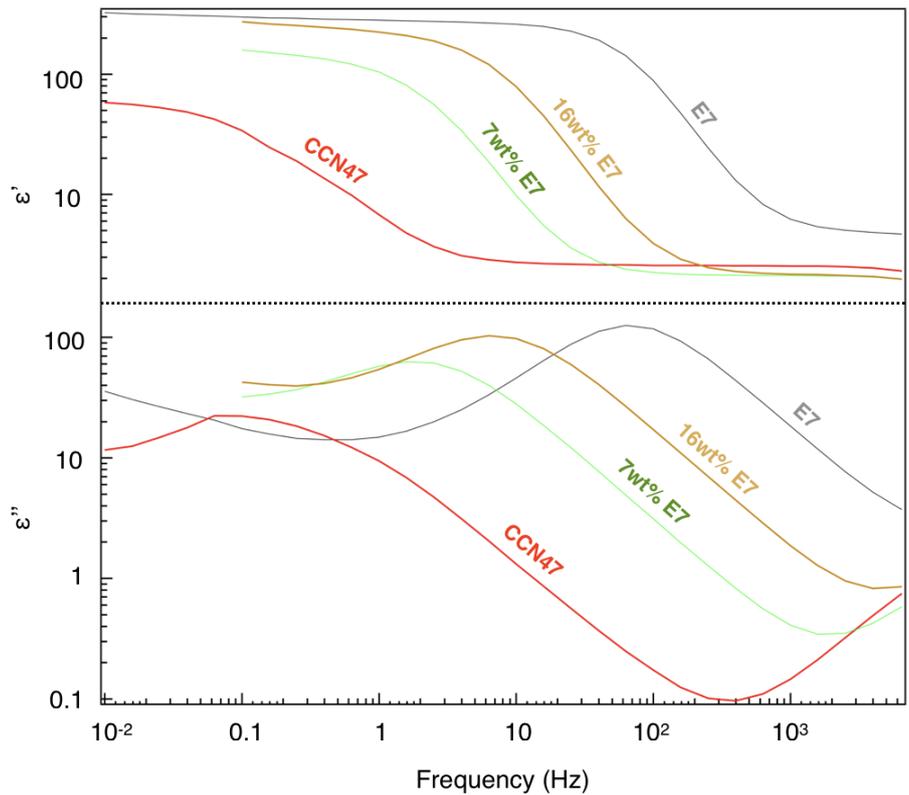

Figure S13. Dielectric spectra of CCN47-E7 mixtures. Top: Real part of the complex dielectric constant as a function of the frequency. Bottom: Imaginary part of the complex dielectric constant as a function of the frequency.

**Supplementary References**

Characterization method of ions in liquid crystal materials by complex dielectric constant measurements. *Jpn. J. Appl. Phys.* **38**, 1423–1427 (1999).